# Global and Local Characterization of Rock Classification by Gabor and DCT Filters with a Color Texture Descriptor

J. Wognin Vangah[1], Sié Ouattara[2], Gbélé Ouattara[3], Alain Clement[4]

URMI Electronique et Electricité Appliquées (EAA)
Institut National Polytechnique Felix Houphouët-Boigny (INP-HB)
BP 1093 Yamoussoukro, Côte d'Ivoire[1, 2, 3]
Laboratoire Angevin de Recherche en Ingénierie des Systèmes (LARIS)
Institut Universitaire de Technologie (IUT) / Université d'Angers
4 Boulevard Lavoisier - BP 42018 - 49016 - ANGERS, France[4]

*Abstract*—In the automatic classification of colored natural textures, the idea of proposing methods that reflect human perception arouses the enthusiasm of researchers in the field of image processing and computer vision. Therefore, the color space and the methods of analysis of color and texture, must be discriminating to correspond to the human vision. Rock images are a typical example of natural images and their analysis is of major importance in the rock industry. In this paper, we combine the statistical (Local Binary Pattern (LBP) with Hue Saturation Value (HSV) and Red Green Blue (RGB) color spaces fusion) and frequency (Gabor filter and Discrete Cosine Transform (DCT)) descriptors named respectively Gabor Adjacent Local Binary Pattern Color Space Fusion (G-ALBPCSF) and DCT Adjacent Local Binary Pattern Color Space Fusion (D-ALBPCSF) for the extraction of visual textural and colorimetric features from direct view images of rocks. The textural images from the two G-ALBPCSF and D-ALBPCSF approaches are evaluated through similarity metrics such as Chi2 and the intersection of histograms that we have adapted to color histograms. The results obtained allowed us to highlight the discrimination of the rock classes. The proposed extraction method provides better classification results for various direct view rock texture images. Then it is validated by a confusion matrix giving a low error rate of 0.8% of classification.

## I. Introduction

In general, the classification and characterization of rocks are done visually following a long process by geologists and mineralogists with many years of experience or through by laboratory tests [1]. Therefore, this so-called manual classification takes time and seems approximate and subjective. However, the automatic classification of rocks could be beneficial and bridge this gap. Today, the analysis and automatic classification of color textures has become one of the areas of active search for shape recognition and computer vision. Several fields of application are covered; let us mention: biomedical [2], facial recognition [3], classification of rocks [4-14]. The automatic classification of rocks is a challenge in the field of image processing with an interest in geologists, universities and specific schools that study rocks and also these applications in construction (roads, buildings, monuments). This is mainly related to the complex nature of the rocks that make those natural textures are not homogeneous, directional natural textures with very different granularity and color properties, making their classification very difficult. In this classification, the physiological perception of texture and color is very important. Therefore, the color space and texture / color analysis methods to be used, must be chosen to match human vision. A good classification always starts from discriminating methods of extraction of texture / color characteristics that are robust to noise, rotation and change of illumination. Typically, the feature extraction process for texture analysis involves statistical, structural, and multiscale methods. However, statistical and multiscale (frequency) approaches for feature extraction are becoming more popular using co-occurrence matrices, histograms, Gabor filters, and various LBP enhancements with exponential use for facial detection [3]. All these characteristics are derived from the measurement of attributes such as energy, contrast, entropy. Then, they are used by various classification and indexing algorithms such as K nearest neighbors (K-NN) [5, 8], Boosting algorithms (LPBoosting) [6], Support Vector Machine (SVM) [7, 9], Artificial Neural Networks (ANN) [8, 10], Maximum Likelihood (MV) [11] and K-means [21] to result in a better classification. In general, the extraction techniques focus on three forms of analysis of visual attribute: spectral analysis, radiometric analysis and textural analysis in the joint or separate use of color and texture. In [10], Ishikawa and Virginia in 2013, based on these visual attributes (texture and color) and Raman spectroscopy, were able to differentiate minerals in igneous rocks from networks of neurons through the analyze of spectral signature of minerals. However, although the classification of minerals has been largely successful, it is difficult to apply the same methods to all rocks because the spectra may have a combination of signature concurrent. In addition, some minerals that were under-represented with their method, were well identified with radiometric and textural analyzes. In [12], Blake et al. (2012), using X-ray diffraction, required rock samples to be collected and pulverized before chemical analysis. In [9], Galdames et al. (2017), based on a textural and colorimetric analysis with 3D Laser Range-based features, made a lithological classification to determine the approximate mineralogical composition of rocks. A method for identifying the texture of different basalts in RGB or grayscale images using neural networks has been introduced by Singh et al. (2010) [13]. In [7], Bianconi et al. (2012) presented a classification system for granite tiles incorporating textural and color features. They tested different characteristics and the best performance was obtained with the co-occurrence matrices. A method for classifying limestone types using features based on color image histograms and a probabilistic neural network has been introduced in [14]. Lepisto et al. (2005) [5], succeeded in classifying rocks with textural and colorimetric analysis of visual attributes by applying Gabor filters in different color channels independently. In this study, it has been shown that it is possible to improve the classification of natural rock images by combining the color information with the description of the texture. It is known that rock images are rich in texture and color information. Despite these cited works, the classification of rock texture images remains a real challenge for the image processing scientific community. In recent times, the texture analysis community has developed a variety of different descriptors for the effective capture of textural information for representation and analysis. Local Binary Pattern (LBP) [15] is one of the best texture descriptors for extracting local texture information and has been used in various applications such as face recognition [3] and rock classification [ 4]. In [4], the performance of LBP, Coordinated Cluster Representation (CCR) and Improved Local Binary Pattern (ILBP) are measured for the classification of granite textures by evaluating the robustness against the rotation of these different LBP descriptors. In 2003, Lepisto et al [16] improved the classification result with the application of K-NN by combining the color information in the HSI space and the extraction of the texture of the rocks by the Gabor filters. The HSI space proved better for this study. Over the past decade, various original LBP extensions have been proposed for classification performance. Attention has been focused on Gabor filters and the LBP operator fusion. Thus Zhang et al. [17] introduced the Local Gabor-based Binary Pattern Histogram Sequence (LGBPHS) by combining LBP and Gabor to enhance the discriminative capacity of LBP descriptors. For the same reasons, Shan et al. [18] proposed Local Gabor Binary Patterns (LGBP) in face recognition. In 2013, Zhihua Xie in [19] showed that integrating global and local characteristics into facial recognition improved performance by comparing the DCT method to the new LGBPH classification method. However, these texture representation methods [17-19] do not consider, on the one hand, the color information in the texture and, on the other hand, its spatial representation in the texture and until now have not yet has been applied to our knowledge to the textures of rock direct view images. To overcome these problems, we propose in this paper two new color texture descriptors that consists in extending LBP color (ALBPCSF) [20] to Gabor Adjacent Local Binary Pattern Color Space Fusion (G-ALBPCSF) by introducing before and impressively the filtering multi-orientation and multi-scale Gabor. And then the DCT Adjacent Local Binary Pattern Color Space Fusion (D-ALBPCSF) compared to the color LBP considering spatial structure relationships and color characteristics. This paper is organized as follows. Section II refers to previous work on LBP and its merger with Gabor. In section III, we will present the proposed new approach. Finally, the experimental results obtained by the two strategies of our new approach are compared to the different methods used. In section IV, our results will be discussed, and then we end with section V with the conclusion and the perspectives.

## II. Previous Work on the LBP and its Merger with Gabor

In this section, a summary of the LBP algorithms is first presented as well as its merger with Gabor.

### A. Brief Review of Original LBP

A common strategy for detecting textures in images is to consider local patches that describe the behavior of the texture around a group of pixels. One of the descriptors that follows this strategy is the LBP descriptor. The LBP operator introduced by Ojala et al. [22] is an effective element of texture description with its discriminating power and simplicity of calculation. As shown in Fig. 1, the operator labels the pixels of an image with decimal numbers, called local bit patterns or LBP codes, which encodes the local structure around each central pixel defined as the threshold for the 3x3 neighborhood. Then each pixel is compared to its eight neighbors by subtracting the value of the central pixel from that of each of these neighbors. The resulting strictly negative values are coded with 0 and the others with 1. Finally, a binary number is obtained by concatenating all the pixels in the direction of the hand of a watch. The decimal value is used for labeling. This process can be expressed mathematically as follows:

$$LBP_{P,R} = \sum_{i=0}^{P-1} s(p_i - p_c) \times 2^i, \quad s(x) = \begin{cases} 1 & if \ x \geq 0 \\ 0 & otherwise \end{cases} \quad (1)$$

where $p_c$ is the gray value of the central pixel, $p_i$ is the gray value of the neighboring pixels, P is the number of neighboring pixels on the circle, R is the radius of the neighborhood circle.

Then primitive extensions considering neighborhoods of different sizes, use of circular neighborhoods, bilinear interpolation of pixel values, and use of uniform patterns are used to estimate pixels that are not exactly in the center pixels: $LBP_{P,R}^{ri}$ and $LBP_{P,R}^{riu2}$ are respectively invariant by rotation of LBP and invariant by uniform rotation of LBP. These two improved LBP operators have been proposed by Ojala et al. in 2002 [15].

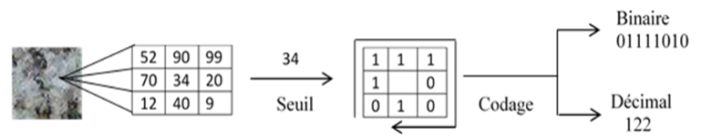

Fig. 1. Example of the Coding Process of the Standard LBP Operator.

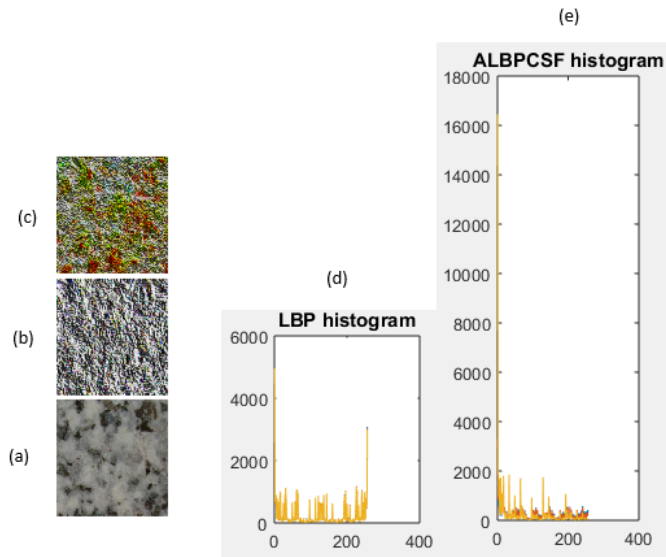

Fig. 2. An Example of an Original Image of Granodiorite (a), Its Image LBP (b), Its Image ALBPCSF (c) (The 3 Superimposed), the Histogram of the Image LBP (d) and the Histogram ALBPCSF (e).

After completing this coding step for the LBP operators, the histogram of these two variants can be created based on the following equation:

$$H_i = \sum_{x,y} I\{f_l(x, y) = i\}, \quad i = 0,\ldots\ldots,n-1$$

*avec*

$$I\{A\} = \begin{cases} 1, & A \text{ is true} \\ 0, & A \text{ is false} \end{cases} \quad (2)$$

where $i$ is the gray level number $i$, $H_i$ is the number of pixels in the image with i as the gray level and n is the number of different labels produced by the LBP operators. Other extensions and variants of the LBP operator followed (see Fig. 2) one another with better performances in the extraction of textures, especially for the classification of facial images but will not be treated in this article. However, an exhaustive review of these LBP color and grayscale descriptors is done in [23, 24] for all readers interested in this descriptor which has gained an exponential popularity over the past ten years [2-4, 9, 15, 17 -20, 23, 24].

### B. Gabor Filter Combined with LBP

To increase the applicability of the LBP operator, modifications of it have been proposed. For example, Zhang et al. [17] proposed the extraction of LBP characteristics from images obtained by filtering a facial image with 40 Gabor filters at different scales and orientations with remarkable results. In [9], the LBP operator was computed on color texture images transformed into the HSV color space after applying Gabor filters on each of the three channels independently. In this same logic, we merge the two RGB and HSV color spaces by putting in the first line the relation between the spatial structure and the color characteristics as in [20].

### III. PROPOSED APPROACHES

Inspired in part by the remarkable achievements obtained by combining Gabor's features with the LBP operator, we propose in this paper two new feature extraction algorithms, the first of which is named Gabor Adjacent Local Binary Pattern color space Fusion (G-ALBPCSF) and the second DCT Adjacent Local Binary Pattern Color Space Fusion (D-ALBPCSF). Indeed, the main interest of the combination of the first strategy, that is G-ALBPCSF compared to the original LBP lies in its ability to model the local characteristics of various orientations and scales provided by the transformation of Gabor and the likely consideration of color information. In G-ALBPCSF, the LBP operator applied on the different channels of the RGB and HSV color spaces, is built on the amplitudes of the Gabor filtered image rather than on the intensity of the original image. This coding makes it possible to exploit multi-resolution information and multi-orientations between the pixels, while being robust to the changes of illuminations. Then the second strategy named D_ALBPCSF, another extraction algorithm that combines the characteristics of the DCT and those of LBPcolor (ALBPCSF) to extract the capital information from the rocks.

### A. Extraction of Characteristics by Gabor Filters

Gabor multi-resolution and multi-scale filters are frequency filters, located in space and with orientations convenient for the extraction and detection of contours. They are applied for decomposing input images for the sequential extraction of characteristics by changing two characteristic parameters that are: frequency and orientation. The Gabor filters for calculating quantities such as amplitude and phase [9] are defined by equation (3):

$$\Psi_{v,\mu}(x, y) = \exp\left(-\frac{|\vec{k}|^2 |\vec{r}|^2}{2\sigma^2}\right) \times \rho$$

with

$$\rho = \left(\exp(i\vec{k}.\vec{r}) - \exp(-\sigma^2/2)\right) \quad (3)$$

$$\vec{r} = \begin{bmatrix} x \\ y \end{bmatrix} \text{ and } \vec{k} = \frac{\pi}{2f^v}\begin{bmatrix}\cos(\mu\pi/8) \\ \sin(\mu\pi/8)\end{bmatrix},$$

$$f = \sqrt{2}, \sigma = \pi$$

wher $\mu = \{0,\ldots,7\}$ and $v = \{0,\ldots,4\}$ define respectively the orientations and the scales of the Gabor filters used.

Gabor's transformation of an image that can be called a Gabor image is defined as the convolution of the original image $I(x, y)$ with the Gabor $\psi_{\mu,v}(x, y)$ filters:

$$G_{\mu,v}(x, y) = I(x, y) * \Psi_{\mu,v}(x, y) \quad (4)$$

The Gabor transformation is a complex function, and can be separated into amplitude $A_{\mu,v}(x, y)$ and phase $\theta_{\mu,v}(x, y)$ and thus can be rewritten as follows:

$$G_{\mu,v}(x, y) = A_{\mu,v} \exp(i\Box\theta_{\mu,v}(x, y)) \quad (5)$$

Equation (5) is a complex representation of the Gabor transformation of the image and from this transformation a feature vector is created (amplitude, phase).

$$A_{\mu,v}(x,y) = |G_{\mu,v}(x,y)|$$

$$\theta_{\mu,v}(x,y) = \arg(G_{\mu,v}(x,y))$$

It should be remembered that since the information phase varies over time (therefore very sensitive), in general, only the amplitude is explored. Thus, for each Gabor filter, an amplitude value is calculated at each position of the pixel, giving a total of 40 amplitudes corresponding to 5 scales and 8 orientations. Fig. 3 shows an image of granite and other images of the same size called Gabor images whose characteristics: amplitude and phase were extracted for two wavelengths and four orientations (16 Gabor images of which 4 are identical so 12 have been represented).

These images clearly show, on the one hand, that the texture information is perceptible and better represented in the Gabor amplitude images, and on the other hand they show that whatever the orientation and / or the scale, the granularity and the structure remain the most important elements and differ significantly from those shown by LBP images (Fig. 3). However, we summarize that Gabor filters transform a given image in only three directions: vertical (0 ° or 180 °), horizontal (90 °) and diagonal (45 ° and 135 °).

### B. Extraction of Characteristics by ALBPCSF

To improve the information in the amplitudes, we code the values of the amplitudes of Gabor by the operator ALBPCSF. In G-ALBPCSF, the ALBPCSF operator is driven on the amplitudes produced by the multi-resolution and multi-scale Gabor filters rather than on the intensities of the original images. The LBP operator uses the comparison between the central pixel and its eight neighbors in a 3 * 3 neighbor and their combination with the Gabor images thus exploiting the links between the pixels for several resolutions and orientations has proved to be very robust to enlightenment and change of scale [17]. Fig. 5 illustrates this combination well. It shows the application of ALBPCSF on five Gabor amplitude images for two different frequencies (λ = 4 for the images of the first line and λ = 8 for the images of the second line) and five orientations. Fig. 4 shows that the Gabor filters are more informative in the diagonal direction (45° and 135°) and confirms the directional structure of the studied rock.

### C. Combination of ALBPCSF with DCT for Extraction of Characteristics

Discrete Cosine Transform (DCT) is a popular technique for imaging and compressing still images and video that transforms spatial representation signals into a frequency representation. As we know, a large amount of information about the original image is stored in a relatively small number of coefficients (in the upper left corresponding to the DCT components at low spatial frequencies of the image) in the image center of Fig. 5. This region contains most of the information, energy, and useful features of the image.

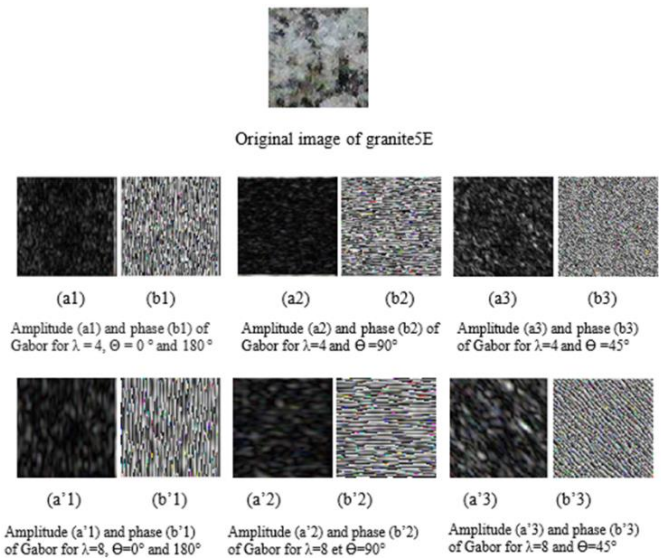

Fig. 3. Examples of an Image Filtered by Four different Gabor Filters: (a1), (a2), (a3), (a'1), (a'2) and (a'3) are the different Images of Amplitudes and (b1), (b2), (b3), (b'1), (b'2) and (b'3) are the Images of the Gabor Phases for Two Wavelengths λ = 4 and λ = 8.

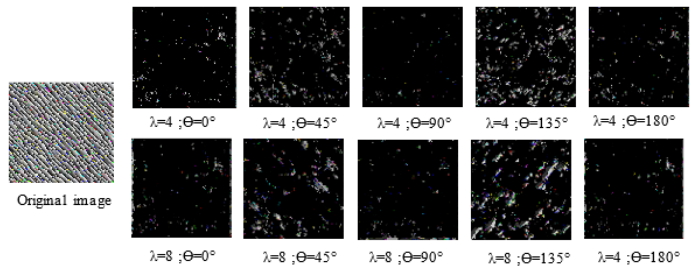

Fig. 4. Gabor_ALBPCSF Image of the Original Granite5E Image for λ = 4 (Top) and λ = 8 (Bottom) and for different Orientations (0 °, 45 °, 90 °, 135 °, 180 °).

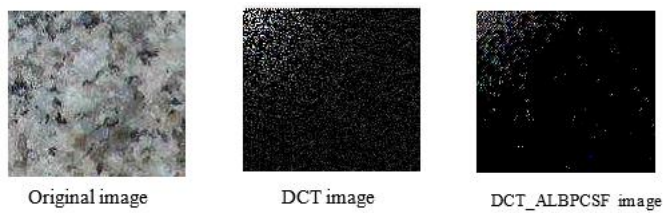

Fig. 5. DCT Image (Middle) of the Original Granite5E Image (Left) and its Combination with ALBPCSF (Right).

For a digital image $f(x,y)$ of resolution $M \times N$, its two-dimensional DCT is defined as follows [19]:

$$A(i,j) = a(i)a(j)\sum_{x=0}^{M-1}\sum_{y=0}^{N-1} f(x,y) \times \cos\left[\frac{(2x+1)i\pi}{2M}\right]\cos\left[\frac{(2y+1)j\pi}{2N}\right]$$

$$i = 0,1,2,.....,M-1; \quad j = 0,1,2,.....,N-1; \tag{6}$$

Where, x and y are the spatial coordinates of the pixels of the image; *i* and *j* are the coordinates of the DCT coefficients of the pixels. $A(i,j)$ is the result of the DCT. $a(i)$ and $a(j)$ is defined as follows:

$$a(i)=\begin{cases}\sqrt{1/M} & \text{if } i=0 \\ \sqrt{2/M} & \text{if } 1\le i\le M-1\end{cases} \quad a(j)=\begin{cases}\sqrt{1/N} & \text{if } j=0 \\ \sqrt{2/N} & \text{if } 1\le j\le N-1\end{cases} \quad (7)$$

However, the performances of the DCT or the LBP taken independently show some insufficient. In [19], the combination of DCT with LGBPH on an ENT database showed better results compared to LGBPH applied alone. In this section, we propose a new algorithm for extracting rock texture features named D-ALBPCSF by combining the DCT and ALBPCSF to consider the color information in addition to the advantages already mentioned above. First, the DCT is performed on direct images of rock. Then we select some useful low frequency DCT coefficients to extract the overall characteristics of the rock texture images. In addition, the LBPcolor operator (more precisely ALBPCSF) is executed on these DCT characteristics of rock textures to now extract the local characteristics with high frequencies (see image on the far right in Fig. 5 above). As we know, the analysis of the existing has shown that using a combination of several classifiers of different types can improve the performance of the classification [5]. Indeed, descriptors based on global characteristics can contribute to a discriminant capacity complementary to descriptors based on local characteristics in the recognition of rocks [9].

## IV. EXPERIMENTAL RESULTS AND DISCUSSION

In this section we will analyze and comment on the results of our experiments. The selected direct view rock images shown in Fig. 6 below are images from our designed database. This database contains 160 images of magmatic and metamorphic rocks textures. These images have been grouped into eight classes (from class 1 to class 8). These 256x256 resolution images, encoded on eight bits by colorimetric component, present textures that are not homogeneous and often show significant differences in directionality, granularity and color within a given class. Thanks to these texture and color characteristics, the original images were subjectively grouped into eight classes by a geology expert.

### A. Averages and Standard Deviation of Gabor and DCT Coefficients

In our study, because of the directionality present in the rock textures, the Gabor filters were applied on these textures. Then, first-order (mean), second-order (standard deviation) were computed on Gabor amplitudes and DCT coefficients, then those of Gabor and DCT combined with ALBPCSF in the different channels. color. Indeed, the calculated average characterizes the luminous intensity of the energies in the image whereas the standard deviation characterizes the variation of the average intensity of all the pixels and corresponds to the changes of contrast of the image. The results are recorded in the tables below.

The analysis of the statistical characteristics used in this work such as average and the standard deviation of the DCT coefficients and Gabor amplitudes for an orientation and a scale (lambda = 4, theta = 135°) indicated in Tables I to VIII shows that magmatic rocks have higher intensity rate than metamorphic rocks. These high intensity rate in magmatic rocks indicate that these rocks have a grainy texture and crystallize at higher temperatures and pressures than metamorphic rocks. Magmatic rocks are characterized by strong energies. However, shale, eclogite, and corneal metamorphic rocks have intensities comparable to those of magmatic rocks when DCT coefficients and Gabor amplitudes are used individually, indicating a lack of such approaches.

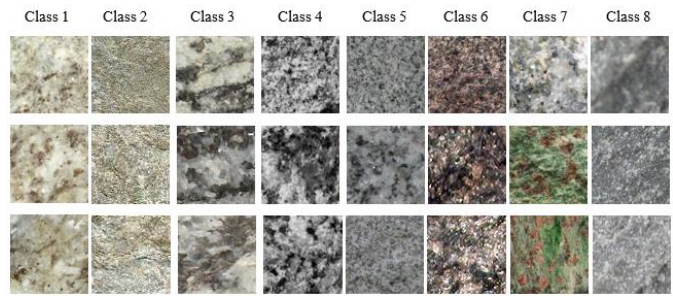

Fig. 6. Examples of the Eight different Classes of Rock Images.

TABLE I. AVERAGE GABOR AMPLITUDE COEFFICIENTS

|  | Average of the coefficients of Gabor amplitudes calculated for different rocks, Lambda = 4, theta = 135° | | | | | | | |
|---|---|---|---|---|---|---|---|---|
|  | Schiste | Gabbro | Granodiorite | Granite | Eclogite | Migmatite | Corneal | Cipolin |
| (R,V) | 65.3018 | 46.945 | 55.5565 | 57.9142 | 44.8483 | 42.2141 | 57.9337 | 42.2962 |
| (G,V) | 65.4266 | 45.9711 | 55.5565 | 57.9107 | 44.7356 | 42.4674 | 57.9511 | 42.8727 |
| (B,V) | 65.4139 | 50.1033 | 55.5565 | 57.9674 | 44.7492 | 42.5286 | 57.9884 | 43.7508 |

TABLE II. AVERAGES OF GABOR_ALBPCSF AMPLITUDE COEFFICIENTS

|  | Mean coefficients of Gabor_ALBPCSF amplitudes calculated for different rocks, Lambda = 4, theta = 135° | | | | | | | |
|---|---|---|---|---|---|---|---|---|
|  | Gabbro | Granodiorite | Granite | Eclogite | Migmatite | Schiste | Corneal | Cipolin |
| (R,V) | 16.2430 | 21.4266 | 15.4365 | 7.1122 | 7.8196 | 6.0078 | 6.0311 | 1.3431 |
| (G,V) | 15.3746 | 21.4266 | 15.4598 | 7.0253 | 7.9517 | 6.0249 | 6.0753 | 1.4434 |
| (B,V) | 18.3144 | 21.4266 | 15.5244 | 6.9742 | 7.9593 | 5.9887 | 6.0438 | 1.4487 |

TABLE III. AVERAGES OF THE DCT COEFFICIENTS

| | Averages of the DCT coefficients calculated for different rocks | | | | | | | |
|---|---|---|---|---|---|---|---|---|
| | Schiste | Gabbro | Granodiorite | Granite | Eclogite | Migmatite | Corneal | Cipolin |
| (R,V) | 13.7518 | 13.2398 | 10.7357 | 8.5716 | 8.8443 | 8.0340 | 6.7993 | 5.6949 |
| (G,V) | 13.7803 | 12.8244 | 10.7357 | 8.5775 | 8.7991 | 8.1157 | 6.8064 | 5.8441 |
| (B,V) | 13.8325 | 13.8325 | 10.7357 | 8.6141 | 8.8009 | 8.1050 | 6.8198 | 5.9921 |

TABLE IV. AVERAGES OF THE DCT_LBP COEFFICIENTS

| | Averages of the DCT_LBP coefficients calculated for different rocks | | | | | | | |
|---|---|---|---|---|---|---|---|---|
| | Gabbro | Granodiorite | Granite | Eclogite | Migmatite | Schiste | Corneal | Cipolin |
| (R,V) | 7.2281 | 5.361 | 1.797 | 1.4901 | 1.1892 | 0.6873 | 0.479 | 0.3811 |
| (G,V) | 6.7981 | 5.361 | 1.7917 | 1.4537 | 1.2444 | 0.6986 | 0.4772 | 0.4472 |
| (B,V) | 7.379 | 5.361 | 1.8362 | 1.4636 | 1.2459 | 0.7399 | 0.4808 | 0.5074 |

TABLE V. STANDARD DEVIATION OF GABOR AMPLITUDE COEFFICIENTS ON ORIGINAL IMAGES

| | Mean standard deviations of Gabor coefficients calculated for different rocks, Lambda = 4, Theta = 135 ° | | | | | | | |
|---|---|---|---|---|---|---|---|---|
| | Schiste | Gabbro | Granodiorite | Granite | Eclogite | Migmatite | Corneal | Cipolin |
| (R, V) | 36.2668 | 28.2244 | 33.5475 | 35.0280 | 27.5958 | 29.0953 | 34.2396 | 28.0670 |
| (G, V) | 36.3339 | 27.9218 | 33.5475 | 35.0924 | 27.6712 | 29.3490 | 34.2445 | 28.6090 |
| (B,V) | 36.3649 | 30.3900 | 33.5475 | 35.0512 | 27.6677 | 29.3802 | 34.2505 | 28.8726 |

TABLE VI. STANDARD DEVIATION OF THE COEFFICIENTS OF AMPLITUDES OF GABOR_ALBPCSF (G-ALBPCSF)

| | Mean standard deviations of G-ALBPCSF coefficients calculated for different rocks, Lambda = 4, Theta = 135 ° | | | | | | | |
|---|---|---|---|---|---|---|---|---|
| | Schiste | Gab bro | Granodiorite | Granite | Eclo gite | Migmatite | Corneal | Cipolin |
| (R, V) | 27.7572 | 44.4104 | 50.9072 | 43.7388 | 28.1538 | 31.2973 | 26.5383 | 12.9097 |
| (G, V) | 27.7616 | 43.1394 | 50.9072 | 43.7826 | 27.9202 | 31.5429 | 26.6805 | 13.3843 |
| (B,V) | 27.6069 | 46.5493 | 50.9072 | 43.8905 | 27.8314 | 31.4801 | 26.6287 | 13.3801 |

TABLE VII. STANDARD DEVIATION OF THE DCT COEFFICIENTS ON THE ORIGINAL IMAGES

| | Mean standard deviations of the DCT coefficients calculated for different rocks | | | | | | | |
|---|---|---|---|---|---|---|---|---|
| | Schiste | Gab bro | Granodiorite | Granite | Eclo gite | Migmatite | Corneal | Cipolin |
| (R, V) | 21.8199 | 27.2926 | 26.2376 | 20.0985 | 20.8450 | 19.4737 | 16.2005 | 15.5525 |
| (G, V) | 21.8907 | 25.7220 | 26.2376 | 20.1102 | 20.6854 | 19.8264 | 16.2257 | 16.3226 |
| (B,V) | 22.1734 | 26.5375 | 26.2376 | 20.3104 | 20.7362 | 19.7837 | 16.2660 | 17.0531 |

TABLE VIII. ECART-TYPE DES COEFFICIENTS DCT_ALBPCSF (D-ALBPCSF)

| | Mean standard deviations of D-ALBPCSF coefficients calculated for different rocks | | | | | | | |
|---|---|---|---|---|---|---|---|---|
| | Schiste | Gab bro | Granodiorite | Granite | Eclo gite | Migmatite | Corneal | Cipolin |
| (R, V) | 7.9382 | 28.0933 | 24.6956 | 13.9308 | 12.2155 | 11.0881 | 6.2382 | 5.9663 |
| (G, V) | 8.0066 | 27.1244 | 24.6956 | 13.9060 | 12.0336 | 11.3821 | 6.4121 | 6.6028 |
| (B,V) | 8.2811 | 28.2027 | 24.6956 | 14.1372 | 12.1098 | 11.3792 | 6.4332 | 7.0632 |

However, the analysis of the means and the standard deviations of the G-ALBPCSF and D-ALBPCSF coefficients calculated from the combinations of the approaches for these same rocks show a good categorization of the rocks:

- Magmatic rocks with higher intensity rate
- Metamorphic rocks with lower intensity rate than those of magmatic rocks.

These analyzes show that the approach combinations D-ALBPCSF and G-ALBPCSF make it possible to better categorize the rocks, unlike the information with Gabor and DCT taken individually. However, the reflection that this raise is up to what threshold value or limit, we can consider that the average characteristics corresponds to that of a rock belonging to a given class?

### B. Comparison D-ALBPCSF and G-ALBPCSF based on the Measurements of Similarity of Intersection of Histograms and chi 2

There are several frequently used metrics for measuring similarity between two histograms for comparing textures. These metrics calculate the distance between the characteristic vectors. In this study, we use the intersection of the histograms and the distance of the Chi 2 defined respectively in equations (8) and (9) below and will allow to compare the efficiency of these two types of characteristics described in Section 3 a little higher.

$$HI(h_i(1), h_i(2)) = 1 - \sum_i \min(h_i(1), h_i(2)) \qquad (8)$$

$$\chi^2(h_i(1), h_i(2)) = \sum_i \frac{(h_i(1) - h_i(2))^2}{(h_i(1) + h_i(2))} \qquad (9)$$

where $h_i$ are the histograms of the samples of the images of the rocks to be compared. Rock samples can be said to be similar by matching their characteristic histograms (color and / or texture) and when the value obtained is closer to 0 if not there is no similarity. The same process is followed for Chi2. The results with both metrics show good trends with very low values with the $\chi 2$. Values between 0 and 1 show that the histograms have been normalized. However, the HI and $\chi 2$ applications have better results in terms of similarity between the rocks with the combination D-ALBPCSF compared to G-ALBPCSF and the other methods used. This shows that the characteristics extracted by the DCT are practically common to all rocks of the same class and show a strong correlation between the rocks of the same class. Tables IX and X show the results of these first experiments.

In this work, we will manly exploit he intersection metric of the histograms with the LBP and D_ALBPCSF approaches given that these approaches make it easier to appreciate the similarity with respect to G_ALBPCSF (see Tables XI and XII).

The analysis of these two methods shows that some metamorphic rocks (migmatite, eclogite, cornea, ...) are like each other and to some magmatic rocks (granite, gabbro, ...). This can be justified by the fact that they have magmatic origins since genesis as it is the case for eclogite for example. The G-ALBPCSF strategy will be especially exploited in our future work. The results are shown in Fig. 7.

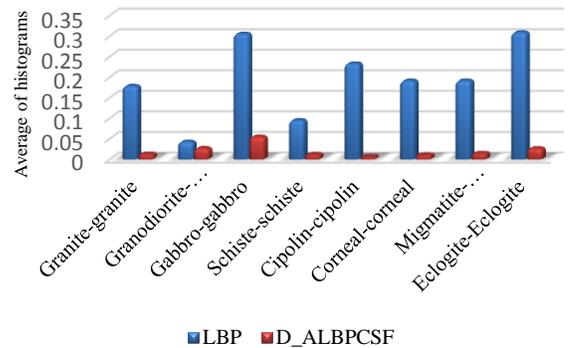

Fig. 7. Intersection of Histograms of Rocks with LBP and D_ALBPCSF.

TABLE IX. MEAN VALUES OF HI FOR THE DIFFERENT METHODS USED

| | Mean HI values of the different methods used | | | | |
|---|---|---|---|---|---|
| Rocks | ImagRGB | LBP | ALBPCSF | DALBPCSF | GALBPCSF |
| Schiste | 0.4375 | 0.0977 | 0.2822 | 0.0122 | 0.2422 |
| Granite | 0.6737 | 0.2188 | 0.3847 | 0.0143 | 0.2665 |
| Ciprolin | 0.6678 | 0.2663 | 0.3945 | 0.0077 | 0.5534 |
| Eclogite | 0.9283 | 0.3756 | 0.5696 | 0.0799 | 0.5673 |
| Granodiorite | 0.0635 | 0.0577 | 0.041 | 0.0359 | 0.0843 |
| Migmatite | 0.5690 | 0.2199 | 0.3805 | 0.0137 | 0.3492 |
| Gabbro | 0.7345 | 0.3358 | 0.4689 | 0.0764 | 0.6444 |
| Corneal | 0.5457 | 0.194 | 0.3680 | 0.0096 | 0.3918 |

TABLE X. AVERAGE VALUES OF χ2 OF THE DIFFERENT METHODS USED

| | Average values of $\chi^2$ of the different methods used | | | | |
|---|---|---|---|---|---|
| Rocks | ImagRGB | LBP | ALBPCSF | DALBPCSF | GALBPCSF |
| Schiste | $25{,}2.10^{-4}$ | $3{,}59.10^{-4}$ | $14{,}5.10^{-4}$ | $0{,}62710^{-4}$ | $10{,}278.10^{-4}$ |
| Granite | $44.10^{-4}$ | $7{,}75.10^{-4}$ | $20{,}4.10^{-4}$ | $0{,}62710^{-4}$ | $11{,}5610^{-4}$ |
| Ciprolin | $53.10^{-4}$ | $13.10^{-4}$ | $14{,}25.10^{-4}$ | $0{,}3229.10^{-4}$ | $33.10^{-4}$ |
| Eclogite | $71.10^{-4}$ | $13{,}5.10^{-4}$ | $36.10^{-4}$ | $1{,}4872.10^{-4}$ | $30{,}25.10^{-4}$ |
| Granodiorite | $1{,}02.10^{-4}$ | $0{,}44.10^{-4}$ | $0{,}220.10^{-4}$ | $0{,}3971.10^{-4}$ | $0{,}9283.10^{-4}$ |
| Migmatite | $33{,}2.10^{-4}$ | $9{,}014.10^{-4}$ | $20{,}6.10^{-4}$ | $0{,}4892.10^{-4}$ | $17.10^{-4}$ |
| Gabbro | $52{,}8.10^{-4}$ | $19.10^{-4}$ | $28{,}6.10^{-4}$ | $2{,}7523.10^{-4}$ | $44{,}8.10^{-4}$ |
| Corneal | $31{,}6.10^{-4}$ | $7{,}47.10^{-4}$ | $19{,}2.10^{-4}$ | $0{,}3398.10^{-4}$ | $20{,}8.10^{-4}$ |

The figure shows a good similarity between the rocks of the same class and the D_ALBPCSF method showing the relevance of our proposed method compared to the existing LBP method.

*C. Matrix of Confusion*

A confusion matrix (Table XIII) was performed with the D_ALBPCSF method discussed in Section 3.2 above, to show the relevance of this method. The effectiveness of the latter is evaluated with the selection of five images of each rock class (8 classes), i.e. 48 images in total for a classification of 320 crossings. For this experiment, the sensitivity (recall), specificity, accuracy and error rate of which the equations are described below and noted respectively (10), (11), (12) and (13) are performance indicators that were chosen to approve the effectiveness of the proposed method and that were also used by Vivek and Audithan in 2014 [6]. Sensitivity is the quality of a class. It indicates the likelihood of a rock to belong to the class knowing that it should belong to it, more simply it is the rate of true positives. Specificity indicates, for its part, the probability that a rock does not belong to its class appropriately, it is the rate of true negatives while the rate of errors corresponds to the general quality of the model.

$$sensitivity = FVP = \frac{VP}{VP + FN} \quad (10)$$

$$specificity = FVN = \frac{VN}{VN + FP} \quad (11)$$

$$precision = \frac{VP}{VP + FP} = \frac{VN}{VN + FN} \quad (12)$$

$$accuracy = \frac{VP + VN}{VP + VN + FP + FN} \quad (13)$$

where VP: True Positive, FP: False Positive; VN: True Negative; FN: False Negative. Here the definitions of the terms assume that one rock is like another if the value of the intersection of their histogram is the lowest. As a result, a true positive is a rock that belongs to a class and whose average value of the histogram intersections is the smallest. The false positive is a rock that does not belong to a class but has the average value of the smallest inter-sections of histograms. The false negative is a rock that does not belong to a class and does not have the smallest average value of the histogram intersections. The true negative is a non-class rock that does not have the smallest value of the histogram intersections.

Both methods give similar performance at first sight from a general analysis of performance indicators such as sensitivity, specificity and accuracy. However, with a misclassification rate of almost 8% with both the LBP and D_ALBPCSF methods for the 5 classes of metamorphic rocks, the proposed method classifies magmatic rocks better with an error rate of 0.8% against 3.3% for the LBP method. These results show a slightly better performance of our method compared to LBP (see Fig. 8). This is consistent with the observation that can be made of these rock texture images. For images of magmatic rocks, we notice that the texture and the color in these images are somewhat regular in their spatial representation, unlike the case of certain metamorphic rocks where there is no homogeneity in the spatial representation of the tex. and color (examples of eclogite and migmatite). The color texture combination in the rock study has been beneficial for their identification.

As a result, the local and global characteristics used in this study for their extraction have been very useful.

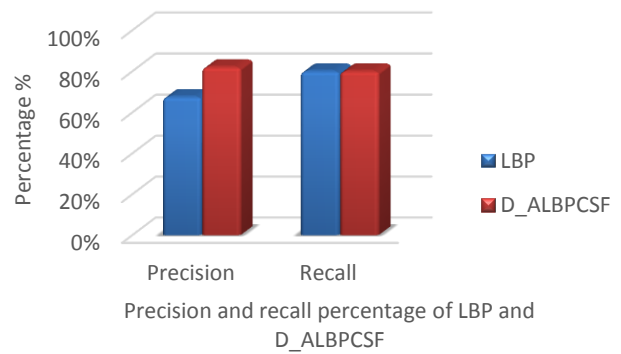

Fig. 8. Precision and Recall of LBP and D_ALBPCSF Methods.

*(IJACSA) International Journal of Advanced Computer Science and Applications,*
*Vol. 10, No. 4, 2019*TABLE XI. CONFUSION MATRIX WITH THE D_ALBPCSF METHOD

| | | Class 1 | Class 2 | Class 3 | Class 4 | Class 5 | Class 6 | Class 7 | Class 8 |
|---|---|---|---|---|---|---|---|---|---|
| REAL CLASSES | Class 1 | 4 | 0 | 0 | 0 | 1 | 0 | 0 | 0 |
| | Class 2 | 0 | 5 | 0 | 0 | 0 | 0 | 0 | 0 |
| | Class 3 | 0 | 0 | 5 | 0 | 0 | 0 | 0 | 0 |
| | Class 4 | 0 | 0 | 0 | 5 | 0 | 0 | 0 | 0 |
| | Class 5 | 0 | 0 | 0 | 0 | 5 | 0 | 0 | 0 |
| | Class 6 | 1 | 1 | 0 | 0 | 0 | 2 | 0 | 1 |
| | Class 7 | 0 | 0 | 0 | 0 | 0 | 0 | 4 | 1 |
| | Class 8 | 1 | 0 | 0 | 0 | 2 | 0 | 0 | 2 |

TABLE XII. PERFORMANCE INDICATORS OF THE TWO METHODS

| Assessment indicators | | D_ALBPCSF | LBP |
|---|---|---|---|
| Images of rocks textures for different classes | VP | 32 | 32 |
| | VN | 272 | 272 |
| | FP | 8 | 8 |
| | FN | 8 | 8 |
| | Sensitivity | 0,8=80% | 0,8=80% |
| | Specificity | 0,97=97% | 0,97=97% |
| | Accuracy | 0,95=95% | 0,95=95% |
| | Error rate | 0,05 = 5% | 0,05 = 5% |

TABLE XIII. COMPARISON OF LBP AND D_ALBPCSF METHODS

| | Parameters of performance | Granite | Granodiorite | Gabbro | Schiste | Cipolin | Eclogite | Migmatite | Corneal |
|---|---|---|---|---|---|---|---|---|---|
| LBP | | VP=5 | VP=5 | VP=5 | VP=5 | VP=5 | VP=0 | VP=5 | VP=2 |
| | | FN=0 | FN=0 | FN=0 | FN=0 | FN=0 | FN=5 | FN=0 | FN=3 |
| | | VN=31 | VN=35 | VN=35 | VN=35 | VN=33 | VN=35 | VN=33 | VN=32 |
| | | FP=4 | FP=0 | FP=0 | FP=0 | FP=2 | FP=0 | FP=2 | FP=3 |
| | Accuracy by class | 90% | 100% | 100% | 100% | 95% | 87,50% | 95% | 85% |
| | Positive accuracy | 55,60% | 100% | 100% | 100% | 71,40% | 0% | 71,40% | 40% |
| | Negative accuracy | 100% | 100% | 100% | 100% | 100% | 87,50% | 100% | 94,10% |
| | Average accuracy | Magmatic rocks : 96,7% | | | Metamorphic rocks : 92,5% | | | | |
| DALBPCSF | | VP=5 | VP=5 | VP=5 | VP=4 | VP=5 | VP=2 | VP=4 | VP=2 |
| | | FN=0 | FN=0 | FN=0 | FN=1 | FN=0 | FN=3 | FN=1 | FN=3 |
| | | VN=34 | VN=35 | VN=35 | VN=33 | VN=32 | VN=35 | VN=35 | VN=32 |
| | | FP=1 | FP=0 | FP=0 | FP=2 | FP=3 | FP=0 | FP=0 | FP=3 |
| | Accuracy by class | 97,50% | 100% | 100% | 92,50% | 92,50% | 92,50% | 97,50% | 85% |
| | Positive accuracy | 83,30% | 100% | 100% | 66,70% | 62,50% | 100% | 100% | 40% |
| | Negative accuracy | 100% | 100% | 100% | 97,10% | 100% | 92,10% | 97,20% | 94,10% |
| | Average accuracy | Magmatic rocks : 99,2% | | | Metamorphic rocks : 92% | | | | |

9 | P a g e
www.ijacsa.thesai.org

## V. Conclusion

In this article, we present two new feature extraction methods applied to rock images. We combined the functionalities of Gabor-ALBPCSF and DCT-ALBPCSF to propose two new descriptors G_ALBPCSF and D_ALBPCSF and to better classify rock texture images. In general, this is a very difficult classification task because of the frequent differences observed within samples of the same type of rock. Experimental results on our direct view rock image database show that the G_ALBPCSF and D_ALBPCSF combinations improve the recognition performance compared to the original LBP and ALBPCSF taken separately. That makes it possible to understand that local and global information must be considered for the extraction of rock characteristics.

In perspective, we plan to apply the K-SVD method to the proposed methods and then apply our methods to other types of images such as facial images.

References

[1] Chatterjee, S., Bhattacherjeeb, A., Samanta, B., Pal, S.K., "Image-based quality monitoring system of limestone ore grades'', Computers in Industry, Vol. 61, no. 5, pp. 391–408, 2010.

[2] F. Bianconi and A. Fernández : An appendix to ''Texture databases – A comprehensive survey'', Pattern Recognition Letters, Vol. 45, pp. 33-38, 2014.

[3] Choi JY, Ro YM, Plataniotis KN, "Color local texture features for color face recognition'', *IEEE Transactions on Image Processing*. Vol. 21, no. 3, pp. 1366–1380, 2012.

[4] A. Fernandez, O. Ghida, E. Gonzalez, F. Bianconi, P.F. Whelan, " Evaluation of robustness against rotation of LBP, CCR and ILBP features in granite texture classification'', Machine Vision and Applications, Vol. 22, pp. 913-926, 2011.

[5] L. Lepisto, I. Kunttu, and A. Visa, " Rock images classification using color features in Gabor space'', Journal of Electronic Imaging (JEI letters), Vol .14, no. 4, 2005.

[6] C. Vivek and S. Audithan, "Robust analysis of the rock texture image based on the boosting classifier with Gabor wavelet features'', Journal of Theoretical and Applied Information Technology, Vol. 69, no. 3, pp. 562-570, 2014.

[7] F. Bianconi, E. Gonzalez, A. Fernández and S. A. Saetta, "Automatic classification of granite tiles through colour and texture features", Expert System with Application Vol. 39, pp. 11212-11218, 2012.

[8] M. Mlynarczuk, M. Skiba, "The Application of Artificial Intelligence for the Identification of the Maceral Groups and Mineral Components of Coal", Computers and Geosciences, Vol. 103, pp. 133-141, 2017.

[9] F. J. Galdames, C. A. Perez, P. A. Estévez and M. Adams, "Rock Lithological Classification by Laser Range 3D and Color Images", Interational Journal of Mineral Processing, 2017.

[10] Sascha. T. Ishikawa and Virginia C. Gulick, "An automated mineral classifier using Raman spectra", Computers & Geosciences, Vol. 54, pp. 259-268, 2013.

[11] P. Paclik, S. Verzakov, and R.P.W. Duin" Improving the Maximum-Likelihood Co-occurrence Classifier : A Study on Classification of Inhomogeneous Rock Images", SCIA, pp. 998–1008, 2005.

[12] D. Blake, "The development of the CheMin XRD/XRF ; refections on building a spacecraft instrument ", In Arospace Conference, IEEE, pp. 1-8, 2012.

[13] N. Singh, T. N. Singh, A. Tiway and K. M. Sarkar, "Textural identification of basaltic rock mass using image processing and neural network," Computer and Geosciences, Vol. 14, pp. 301- 310, 2010.

[14] A. K. Patel and S. Chatterjee, "Computer vision-based limestone rock-type classification using probabilistic neural network, " Geoscience Frontiers 7, pp. 53-60, 2016.

[15] T. Ojala, M. Pietikainen and T. Maenpaa, "Multiresolution gray-scale and rotation invariant texture classification with local binary patterns, "IEEE Transactions on Pattern Analysis and Machine Intelligence, vol. 24, no. 7, pp. 971-987, 2002.

[16] L. Lepisto, I. Kunttu, J. Autio and A. Visa, " Classification Method for Colored Natural Textures Using Gabor filtering'', IEEE, 12 th International Conference on Image Analysis and Processing 397- 401, 2003.

[17] W. Zhang, S. Shan, W. Gao, X. Chen and H. Zhang, " Local Gabor Binary Pattern Histogram Sequence (LGBPHS) : A novel Non-Statistical Model for Face Representation and Recognition'', Proceedings of the Tenth IEEE International Conference on Computer Vision (ICCV05), 2005.

[18] S. Shan, W. Zhang, Y. Su, X. Chen and W. Gao " Ensemble of Piecewise FDA Based on Spatial Histograms of Local (Gabor) Binary Pattern for Face Recognition'', IEEE 2006, The 18th International Conference on Pattern Recognition (ICPR'06), 2006.

[19] Zhihua Xie, "Single Sample Face Recognition Based on DCT and Local Gabor Binary Pattern Histogram'', Springer, ICIC 2013, pp.435- 442, 2013.

[20] J. Wognin VANGAH, Sié OUATTARA, Alain CLEMENT and Gbélé OUATTARA, "Combination of dictionary learning by k-svd and a colorimetric texture descriptor for improved identification of geological structures : case of rocks'', International Journal of Innovation and Applied Studies (IJIAS), Vol. 24, no. 3, pp. 1193-1208, 2018.

[21] O. Baklanova and O. Shvets, "Methods and Algorithms of Cluster Analysis in the Mining Industry - Solution of Tasks for Mineral Rocks Recognition", In Proceedings of the 11th International Conference on Signal Processing and Multimedia Applications (SIGMAP-2014), pp. 165-171, 2014.

[22] Ojala, T., Pietikäinen, M., Harwood, D. : A comparative study of texture measures with classification based on feature distributions. Pattern Recognition. Vol. 29, pp. 51–59, 1996.

[23] L. Nanni, A. Lumini and S. Brahnam, "Survey on LBP based texture descriptors for classification", Expert Systems with Applications, Vol. 39, pp. 3634-3641, 2012.

[24] S. Banerji, A. Verma and C. Liu, "Novel color LBP descriptors for scene and image texture classification'', in Proceedings of the 15th International Conference on Image Processing, Computer Vision, and Pattern Recognition, Las Vegas, Nevada, pp. 537–543, 18-21 July 2011.